# Solitons in Ideal Optical Fibers – A Numerical Development

*Sólitons em Fibras Óticas Ideais – Um Desenvolvimento Numérico*

Eliandro Rodrigues Cirilo[1]; Paulo Laerte Natti[2]; Neyva Maria Lopes Romeiro[3]; Érica Regina Takano Natti[4]; Camila Fogaça de Oliveira[5]

[1] Professor in the Mathematics Departament at Universidade Estadual de Londrina; ercirilo@uel.br
[2] Professor in the Mathematics Departament at Universidade Estadual de Londrina; plnatti@uel.br
[3] Professor in the Mathematics Departament at Universidade Estadual de Londrina; nromeiro@uel.br
[4] Professor in the Pontifícia Universidade Católica do Paraná - Londrina; erica.natti@pucpr.br
[5] Graduated in Mathematics from Universidade Estadual de Londrina; ca_fogaca@yahoo.com.br

**Abstract**

This work developed a numerical procedure for a system of partial differential equations (PDEs) describing the propagation of solitons in ideal optical fibers. The validation of the procedure was implemented from the numerical comparison between the known analytical solutions of the PDEs system and those obtained by using the numerical procedure developed. It was discovered that the procedure, based on the finite difference method and relaxation Gauss-Seidel method, was adequate in describing the propagation of soliton waves in ideals optical fibers.
**Key words:** Optical communication. Solitons. Finite differences. Relaxation Gauss-Seidel method.

**Resumo**

Este trabalho desenvolveu um procedimento numérico para um sistema de equações diferenciais parciais (EDP's) que descreve a propagação de sólitons em fibras óticas ideais. A validação do procedimento foi implementada a partir da comparação numérica entre as soluções analíticas conhecidas do sistema de EDP's e aquelas obtidas por meio do procedimento numérico desenvolvido. Verificou-se que o procedimento, baseado no método das diferenças finitas e no método de Gauss-Seidel com relaxação, mostrou-se adequado na descrição da propagação das ondas sólitons em fibras óticas ideais.
**Palavras-chave:** Comunicação ótica. Sólitons. Diferenças finitas. Método de Gauss-Seidel com relaxação.

## 1 Introduction

In the last decades, several experiments were carried out aiming at trying to improve the capacities of the optical communication systems. The important issue is how to compensate the dispersion and the nonlinearities in communication systems at long distances (thousands of kilometers) or in high debt ground systems. A good technique that allows simultaneous compensation of such effects had already been proposed by Hasegawa and Tappert, in 1973, though only after the appearance of the optical amplifier could it be applied to practical systems (HASEGAWA; TAPPERT, 1973). This technique is based on the use of optical pulses, whose electrical field has the shape of a hyperbolic secant with some milliwatts of peak potency, and in the compensation of the dispersion by the optical fiber nonlinearities. Such pulses, called solitons, are capable of self-propagation, keeping their shape unchanged in a dispersive and non-linear environment, like the optical fiber. (EILENBERGER, 1981; TAYLOR, 1992).

In the 1980's, the experimental development of communication systems based on optical solitons started. Mollenauer, Stolen and Gordon, in 1980, conducted the first experimental observation of the bright soliton propagation in optical fibers. Hasegawa, in 1984, proposed that optical solitons could be used in long distance communication without the need of repeating stations, including overseas communications (HASEGAWA, 1984; PILIPETSKII, 2006). Since then, several experiments were conducted with the objective of improving the transmission capacity of solitons in optical fibers. Emplit et al., in 1987, carried out the first experimental observation of the dark soliton propagation in optical fibers (EMPLIT et al., 1987). Mollenauer and Smith, in 1988, transmitted soliton pulses over 4,000 kilometers using a phenomenon called the Raman Effect to provide optical gain in the fiber (MOLLENAUER; SMITH, 1988). In 1991, a Bell Labs research team transmitted bright solitons error-free at 2.5 gigabits per second, over more than 14,000 kilometers, using erbium optical fiber amplifiers (MOLLENAUER et al., 1991). In 1998, Thierry Georges and his team at France Telecom, combining optical solitons of different wavelengths, demonstrated data transmission of 1 terabit per second (LE GUEN et al., 1999). In 2000, the practical use of solitons turned into reality when Algety Telecom, then located in Lannion, France, developed undersea telecommunication equipments for the transmission of optical solitons. In this time, intense research was also conducted on the possibility of using vector soliton in optical communications. The use of vector solitons in a birefringence fiber was predicted by Menyuk (MENYUK, 1997) and observed recently (ZHANG et al., 2008; TANG et al., 2008).

In this context of optical communication via solitons, an increase in the number of published works, with the aim of overcoming the several problems that have been found and improving the already proposed methods has been verified. Such theoretical and experimental studies approach themes related to the soliton generation processes (MALOMED et al., 2005; KUROKAWA; TAJIMA; NAKAJIMA, 2007), soliton propagation processes (HASEGAWA, 2000; LATAS; FERREIRA, 2007; TSARAF; MALOMED, 2009) and soliton stability processes (CHEN; ATAI, 1998; DRIBEN; MALOMED, 2007) in optical fibers.

In this work, only the results about scalar bright solitons, named simply as solitons from now on, will be discussed. The study of propagation and stability of femtosecond optical solitons in fibers is affected by several disturbing processes. Usually, the most important ones are group velocity dispersion and optical Kerr effect (intensity dependence of the refractive index). Taking only these into account, the pulse propagation is a soliton described by a system of coupled nonlinear Schrödinger differential equations (HASEGAWA; TAPERT, 1973).

To describe real-world fiber-optic system, it is more realistic to include further effects like power loss or Rayleigh scattering (BÖHM; MITSCHKE, 2007), high-order dispersion and high-order nonlinearities (AGRAWAL, 1995), soliton self-steepening, Raman effect and self-frequency shift (LATAS; FERREIRA, 2007), polarization-mode dispersion (DRIBEN; MALOMED, 2007), nonlinear phase noise (LAU; KAHN, 2007), defects or mends in the optical fiber (TSARAF; MALOMED, 2009), among others. It should be observed that the perturbed coupled nonlinear Schrödinger differential equations systems, that describe wave propagation in real optical fibers, do not present analytical solution known in the literature.

In this work, aiming to study the propagation and the stability of such waves, called quasi-solitons, in real optical fibers, firstly a general numerical procedure is developed for the propagation of solitons in ideal optical fibers. It should be noted that, in the case of ideal fibers, the analytical solution of the problem is known, and this will allow the validation of the numerical procedure.

In the literature there are several numerical approaches whose objective is to describe the propagation of solitons in dielectrical environments, most of which use the finite difference method (ISMAIL, 2004; WANG, 2005; CHEN, MALOMED, 2009), the finite element method (DAG, 1999; ISMAIL, 2008) and the split-step method (LIU, 2009). On the other hand, to solve numerically the resulting system of equations, the authors use various methods like Newton's method (ISMAIL, 2008), Crank-Nicolson method (CHEN, MALOMED, 2009), Runge-Kutta Method (REICH, 2000), among others. A review of the several numerical procedures applied to describe the propagation of solitons in optical fibers is found in DEHGHAN and TALEEI (2010).

In this paper, to describe the propagation of solitons waves in ideal optical fibers, a procedure based on the finite difference method and relaxation Gauss-Seidel method is used. Section 2 presents the soliton analytical solutions for the coupled nonlinear Schrödinger differential equations system that describes the propagation of waves in ideal optical fibers. In the sequence, the characteristics of the soliton wave are correlated with the dielectrical properties of the ideal optical fibers. In Section 3, a numerical procedure for this ideal PDE system is developed. In Section 4, by comparing the obtained numerical results with the known analytical results, the consistency of the developed numerical procedure is verified. In Section 5, the main results of this work are presented.

## 2 Solitons in $\chi^{(2)}$ dielectric fibers

This section studies the coupled non-linear complex PDE system, obtained from Maxwell's equations, which describe the longitudinal propagation of two electromagnetic waves (fundamental and second harmonic modes) in $\chi^{(2)}$ dielectrical optical fibers (AGRAWAL, 1995). The detailed mathematical modeling of this PDE system can be found in GALLEAS et al., 2003. This coupled nonlinear differential equations system is given by

$$I\frac{\partial a_1}{\partial \xi} - \frac{r}{2}\frac{\partial^2 a_1}{\partial s^2} + a_1^* a_2 \exp(-I\beta\xi) = 0$$

$$I\frac{\partial a_2}{\partial \xi} - I\delta\frac{\partial a_2}{\partial s} - \frac{\alpha}{2}\frac{\partial^2 a_2}{\partial s^2} + a_1^2 \exp(I\beta\xi) = 0 ,$$

(1)

where $I = \sqrt{-1}$ is the imaginary unit, $a_1(\xi, s)$ and $a_2(\xi, s)$ are complex variables that represent the normalized amplitudes of the electrical fields of the fundamental and second harmonic waves, respectively, with $a_1^*(\xi, s)$ and $a_2^*(\xi, s)$ as their complex conjugates. The independent variable $s$ has spatial dimension character, whereas the independent variable $\xi$ has temporal character.

The real parameters $\alpha$, $\beta$, $\delta$ and $r$, in (1), are related with the dielectrical properties of the optical fiber and should be adjusted so that the existence of solutions is possible (YMAI et al., 2004). In the limit $|\beta| \to \infty$, the coupled differential equations (1) uncouple in Schrödinger's non-linear equation (MENYUK; SCHIEK; TORNER, 1994). Thus, the $\beta$ quantity is a measure of the generation rate of the second harmonic. The $\alpha$ quantity measures the relative dispersion of the group velocity dispersion (GVD) of fundamental and second harmonic waves in the optical fiber. For values $|\alpha| > 1$, the second harmonic wave has higher dispersion than the fundamental wave and for values $|\alpha| < 1$, it is the fundamental wave that has higher dispersion. The $r$ quantity is the signal of the fundamental GVD wave. When $r = +1$, the fundamental wave is in normal dispersion regime, but if $r = -1$, the fundamental wave is in the anomalous dispersion regime. Finally, the parameter $\delta$ measures the difference of group velocities of fundamental and second harmonic waves, so it accounts for the presence of Poynting vector walk-off that occurs in birefringent media, when propagation is not along the crystal optical axes. Notice that it is possible to choose the characteristics (velocity, width, amplitude, stability, etc.) of the wave to be propagated in the optical fiber, selecting or proposing materials with the appropriate $\alpha$, $\beta$, $\delta$ and $r$ dielectrical properties. In (GALLEAS et al., 2003) a detailed description of the interpretation of such dielectrical quantities is given, relating them with the fiber optical properties.

The PDE system (1) presents solitons solutions (GALLEAS et al., 2003), given by

$$a_1 = \pm \frac{3}{2(\alpha - 2r)} \sqrt{\alpha r \left( \frac{\delta^2}{2\alpha - r} + \beta \right)} \times$$

$$\text{sech}^2 \left[ \pm \sqrt{\frac{1}{2(2r - \alpha)} \left( \frac{\delta^2}{2\alpha - r} + \beta \right)} \left( s - \frac{r\delta}{2\alpha - r} \xi \right) \right] \times$$

$$\exp \left\{ I \left[ \frac{r\delta^2(4r - 5\alpha)}{2(2\alpha - r)^2(2r - \alpha)} - \frac{r\beta}{2r - \alpha} \right] \xi - \frac{I\delta}{2\alpha - r} s \right\} \quad (2)$$

$$a_2 = \frac{3r}{2(\alpha - 2r)} \left[ \frac{\delta^2}{2\alpha - r} + \beta \right] \times$$

$$\text{sech}^2 \left[ \pm \sqrt{\frac{1}{2(2r - \alpha)} \left( \frac{\delta^2}{2\alpha - r} + \beta \right)} \left( s - \frac{r\delta}{2\alpha - r} \xi \right) \right] \times$$

$$\exp \left\{ 2I \left[ \frac{r\delta^2(4r - 5\alpha)}{2(2\alpha - r)^2(2r - \alpha)} - \frac{r\beta}{2r - \alpha} + \frac{\beta}{2} \right] \xi - \frac{2I\delta}{2\alpha - r} s \right\}. \quad (3)$$

In (QUEIROZ et al., 2006), a adapted numerical procedure provided the numerical solution of (1), when $\delta = 0$. In the specific case of the propagation of solitons in special optical fibers, and in optimal situations, the walk-off wave phenomenon can be disregarded (ARTIGAS; TORNER; AKHMEDIEV, 1999), which justifies taking $\delta = 0$ in such situations. However, in the case of non-ideal optical fibers, situation in which necessarily $\delta \neq 0$, a general numerical procedure to the system (1) should be developed. This numerical development is presented in the next section.

## 3 Numerical model for the propagation of solitons in optical fibers

The numerical scheme developed in this work to solve the PDEs system (1), consists in approximating the derivates by finite differences and resolving the algebraic system resulting from the discretization, implicitly, by means of the relaxation Gauss-Seidel method (SMITH, 1990; SPERANDIO, MENDES, MONKEN, 2003).

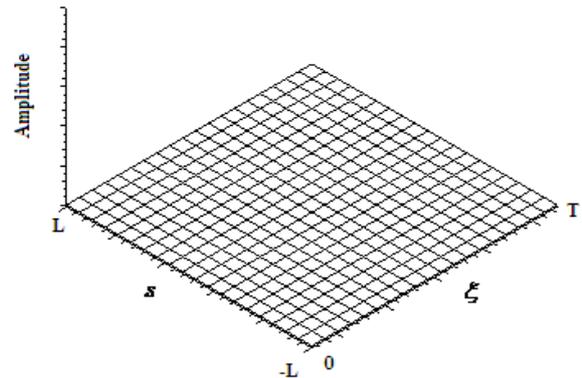

**Figure 1:** Computational domain of the propagation of the soliton waves.

System (1) is numerically resolved in domain $\xi \times s = [0, T] \times [-L, L]$, where $T, L \in \Re$. By discretizing the variables $a_1(\xi, s) \equiv a_1(k+1, i)$ and $a_2(\xi, s) \equiv a_2(k+1, i)$, for $k = 0, 1, ..., k_{max}$ and $i = 1, 2, ..., ni$, where $k_{max}$ is denominated the last advance in $\xi$ and $ni$ the maximum number of points in $s$, the propagation domain of the solitons wave becomes defined by a discretized computational network of $k_{max} \times ni$ points, as represented in Figure 1.

Thus, by means of the method of finite differences, approaching the temporal derivates by progressive differences, and the spatial derivates by central differences (SMITH, 1990), the following linear systems are generated from the differential equations (1), namely,

$$a_1(k+1,i) = \left(\frac{1}{{}^1A_p}\right)\left[{}^1A_W a_1(k+1,i-1) + {}^1A_E a_1(k+1,i+1) + {}^1A_{po} a_1(k,i) - a_1^*(k,i) a_2(k,i) \exp(-I\beta\xi)\right]$$

(4)

$$a_2(k+1,i) = \left(\frac{1}{{}^2A_p}\right)\left[{}^2A_W a_2(k+1,i-1) + {}^2A_E a_2(k+1,i+1) + {}^2A_{po} a_2(k,i) - (a_1(k+1,i))^2 \exp(I\beta\xi)\right],$$

where

$${}^1A_p = \frac{I}{\Delta\xi} + \frac{r}{(\Delta s)^2} \qquad {}^1A_E = {}^1A_W$$

$${}^1A_W = \frac{r}{2(\Delta s)^2} \qquad {}^1A_{po} = \frac{I}{\Delta\xi}$$

$${}^2A_p = \frac{I}{\Delta\xi} + \frac{\alpha}{(\Delta s)^2} \qquad {}^2A_E = \frac{I\delta}{2\Delta s} + \frac{\alpha}{2(\Delta s)^2}$$

$${}^2A_W = -\frac{I\delta}{2\Delta s} + \frac{\alpha}{2(\Delta s)^2} \qquad {}^2A_{p_0} = \frac{I}{\Delta\xi}.$$

In this work, the linear system (4) is resolved by means of the relaxation Gauss-Seidel method (SPERANDIO, MENDES, MONKEN, 2003). Consider this linear system for $a_1(k+1, i)$, given explicitly by

$$a_1(k+1,2) = \left(\frac{1}{{}^1A_p}\right)\left[{}^1A_W a_1(k+1,1) + {}^1A_E a_1(k+1,3) + {}^1A_{po} a_1(k,2) - a_1^*(k,2) a_2(k,2) \exp(-I\beta\xi)\right]$$

$$a_1(k+1,3) = \left(\frac{1}{{}^1A_p}\right)\left[{}^1A_W a_1(k+1,2) + {}^1A_E a_1(k+1,4) + {}^1A_{po} a_1(k,3) - a_1^*(k,3) a_2(k,3) \exp(-I\beta\xi)\right]$$

$$a_1(k+1,4) = \left(\frac{1}{{}^1A_p}\right)\left[{}^1A_W a_1(k+1,3) + {}^1A_E a_1(k+1,5) + {}^1A_{po} a_1(k,4) - a_1^*(k,4) a_2(k,4) \exp(-I\beta\xi)\right]$$

...

$$a_1(k+1,ni-1) = \left(\frac{1}{{}^1A_p}\right)\left[{}^1A_W a_1(k+1,ni-2) + {}^1A_E a_1(k+1,ni) + {}^1A_{po} a_1(k,ni-1) - a_1^*(k,ni-1) a_2(k,ni-1) \exp(-I\beta\xi)\right].$$

It can be written in compact form as

$$a_1(k+1,i) = \left(\frac{1}{{}^1A_p}\right)\left[{}^1B_i + {}^1A_W a_1(k+1,i-1) + {}^1A_E a_1(k+1,i+1)\right],$$

where ${}^1B_i = {}^1A_{po} a_1(k,i) - a_1^*(k,i) a_2(k,i) \exp(-I\beta\xi)$ with $i = 2, ..., ni-1$.

From the initial condition $a_1(0, i)$, given by soliton solution (2), and imposing the contour conditions $a_1(k+1, 1) = 0$ and $a_1(k+1, ni) = 0$, for $L$ sufficiently large, $a_1(k+1, i)^{(n+1)}$ is iteratively calculated by means of the equations

$$a_1(k+1,i)^{(n+1)} = \frac{{}^1B_i^{(n+1)} + {}^1A_W a_1(k+1,i-1)^{(n+1)} + {}^1A_E a_1(k+1,i+1)^{(n)}}{{}^1A_p}$$

(5)

until the stop criterion is fulfilled, namely,

$$\max_{2 \leq i \leq ni-1} |a_1(k+1,i)^{(n+1)} - a_1(k+1,i)^{(n)}| < 10^{-6},$$

(6)

where

$${}^1B_i^{(n+1)} = {}^1A_{po} a_1(k,i)^{(n+1)} - a_1^*(k,i)^{(n+1)} a_2(k,i)^{(n+1)} \exp(-I\beta\xi).$$

This method consists in determining $a_1(k+1, i)^{(n+1)}$ by using the already known components of $a_1(k+1, i+1)^{(n)}$ and $a_1(k+1, i-1)^{(n+1)}$, with the advantage of not requiring the simultaneous storage of the two vectors $a_1(k+1, i+1)^{(n)}$ and $a_1(k+1, i-1)^{(n+1)}$ at each step. Likewise, $a_2(k+1, i)^{(n+1)}$ is resolved.

Notice that, in equations (5)-(6), the value $\omega = 1.0$ was used for the parameter of relaxation (CIRILO et al., 2008). Such value corresponds to the optimal relaxation parameter in relation to the variations of the dielectrical parameters $\alpha$, $\beta$ and $\delta$ of system (1). Figure 2 presents the flowchart of the numerical code developed for the PDEs system (1).

## 4 Numerical Results

This section analyses the numerical code developed in the previous section, comparing the obtained numerical result with the known analytical result (2-3), in function of the dielectrical parameters $\alpha$, $\beta$ and $\delta$ of the studied model.

In the simulations, a CPU with processor AMD Athlon 64X2 Dual Core Processor 4400+, with 2.29 GHz and 1.00 GB of RAM memory is used. In the calculations a tolerance factor of $1.0 \times 10^{-6}$ was imposed. The code was developed in FORTRAN.

For the simulations, the values $r = -1$, $\beta = -1/2$, $\alpha = -1/4$ and $-1/2 < \delta < 1/2$ were adopted. Such values are compatible with experimental measurements observed in commercial optical fibers.

Initially $\delta = 0$ is considered. When $L = 50$ and $T = 10$, from a discretization of 500 points in interval $[-L, L]$ and of $\Delta\xi = 1.0 \times 10^{-2}$, a significant agreement is verified between the numerical and the analytical solutions, as observed in Figures 3 and 4. For all the computational domain, it was obtained that the biggest difference between the analytical and numerical solutions, for the fundamental harmonic, was $1.5 \times 10^{-3}$. Likewise, the maximum numerical error for the second harmonic was $2.7 \times 10^{-3}$. The total processing time was $1.5 \times 10^{-1}$ seconds.

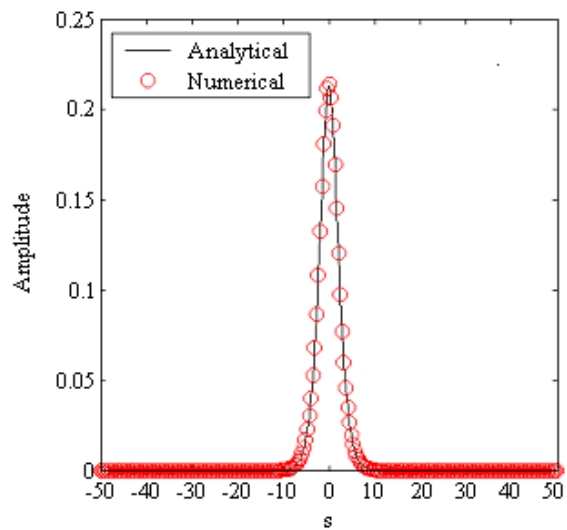

**Figure 3:** Analytical and numerical solutions of $|a_1(\xi, s)|$, in $\xi = 10$, for $\delta = 0$.

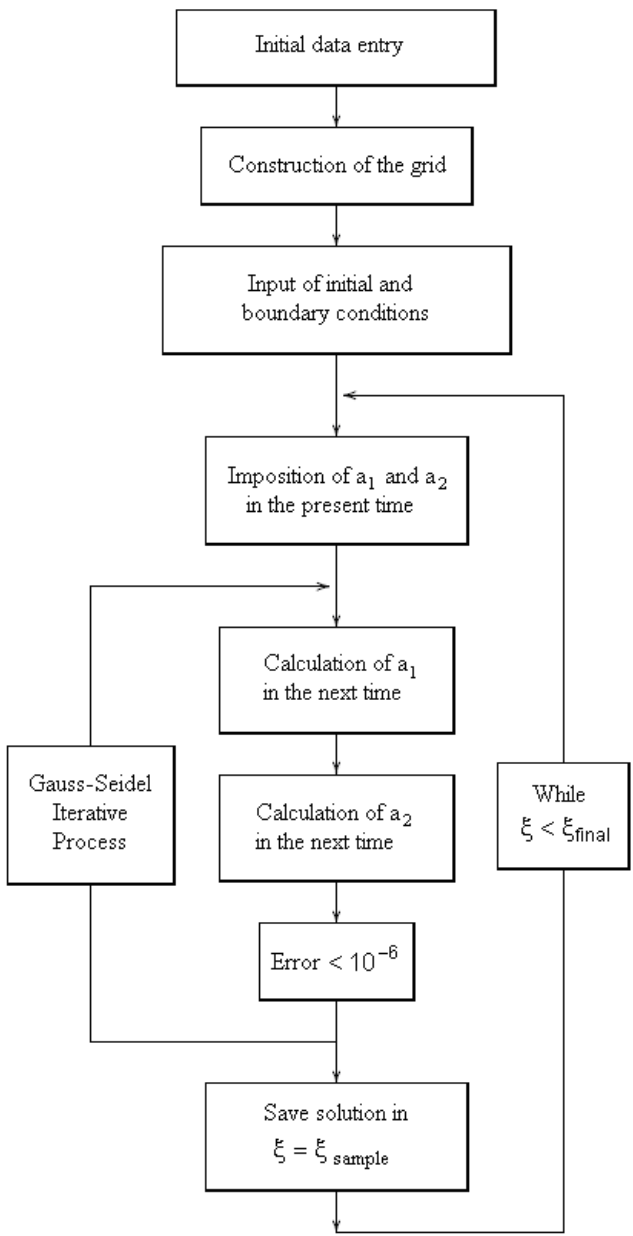

**Figure 2:** Flowchart of the numerical code developed to obtain the numerical solitons solutions.

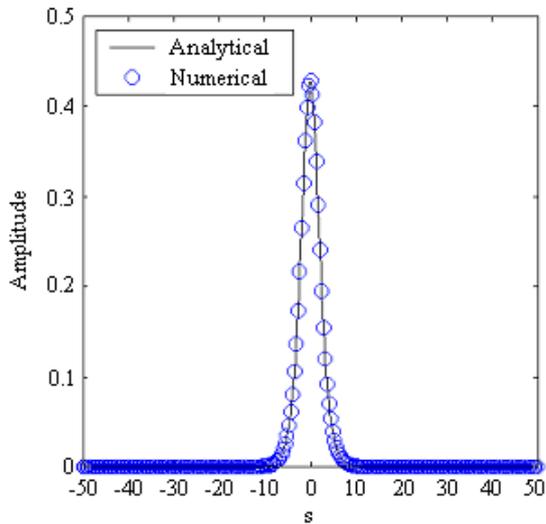

**Figure 4:** Analytical and numerical solutions of $|a_2(\xi,s)|$, in $\xi = 10$, for $\delta = 0$.

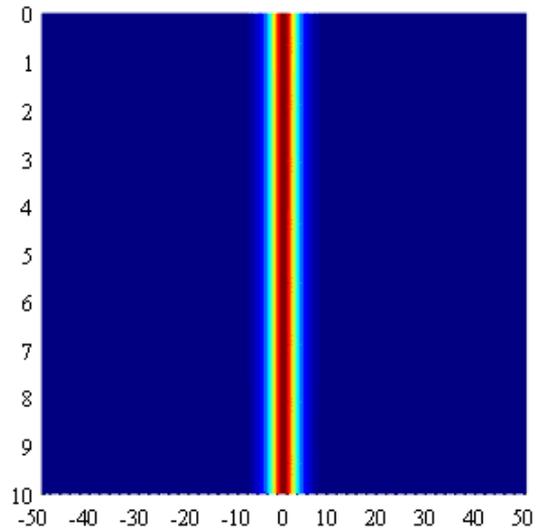

**Figure 6:** Upper view of $|a_1(\xi,s)|$ for $\delta = 0$.

The profile of the fundamental harmonic module, numerically obtained, is presented in Figure 5. From an upper view of plan $\xi \times s$, see Figure 6, it is observed that the soliton remains static in $s = 0$, for all $\xi$. Likewise, the profile of the second harmonic module, also numerically obtained, can be observed in Figure 7, whose upper view is shown in Figure 8. These numerical results are consistent with the analytical results foreseen by (GALLEAS et al., 2003; YMAI et AL., 2004).

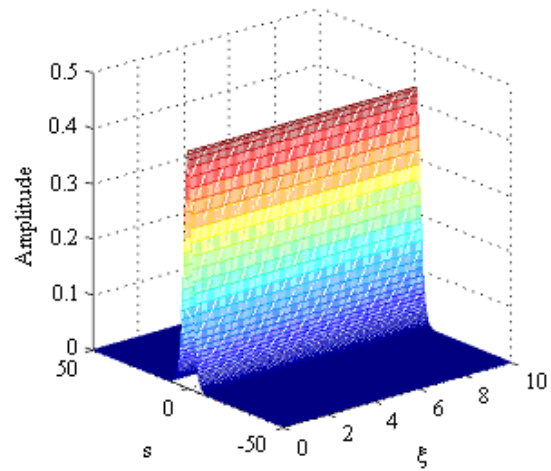

**Figure 7:** Numerical solution of $|a_2(\xi,s)|$, in all computational domain, for $\delta = 0$.

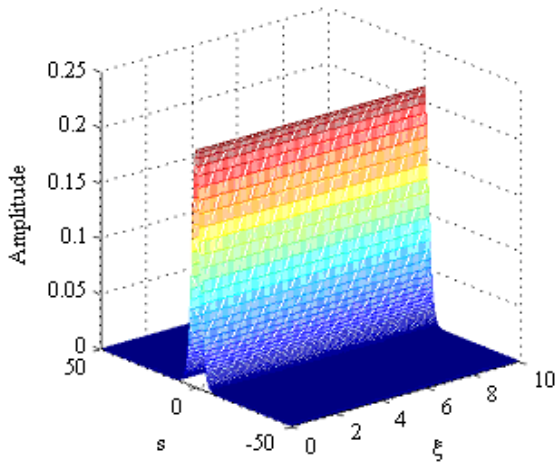

**Figure 5:** Numerical solution of $|a_1(\xi,s)|$, in all computational domain, for $\delta = 0$.

Now the situation $\delta \neq 0$ is considered. Taking $\delta = -1/4$ and the same partition of 500 points in interval $[-L, L]$ and $\Delta\xi = 1.0 \times 10^{-2}$, the propagation of the soliton wave along the fiber is analyzed when $L = 50$ and $T = 10$. The simulations conducted showed that the biggest difference between the analytical and numerical solutions, for the fundamental harmonic, was $7.1 \times 10^{-3}$. Likewise, for the second harmonic, the maximum error was $7.5 \times 10^{-3}$. The total processing time was $1.4 \times 10^{-1}$ seconds. Again, a relevant agreement is observed between these solutions, as observed in Figures 9 and 10.

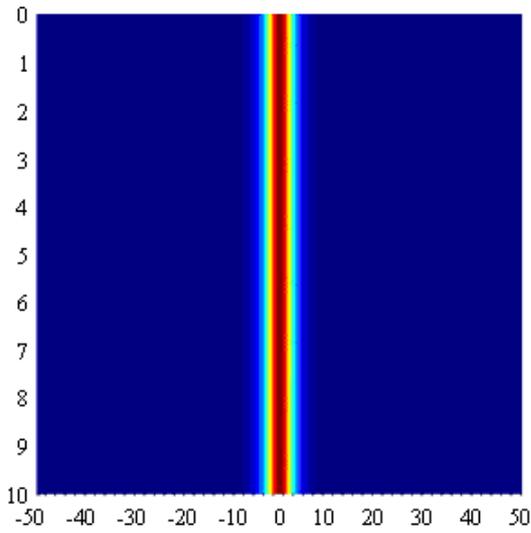

**Figure 8:** Upper view of $|a_2(\xi,s)|$ for $\delta = 0$.

Notice that the maximum errors between the numerical and analytical solutions for $\delta = -1/4$ were slightly superior to the ones obtained for $\delta = 0$. This fact is due to the term $\delta \frac{\partial a_2}{\partial s}$ in (1), which is taken into account in the numerical calculations when $\delta \neq 0$. In other words, the presence of this term in the numerical procedure generates new errors, due to the approximations, which propagate along the computational domain, incrementing the maximum errors between the numerical and analytical solutions.

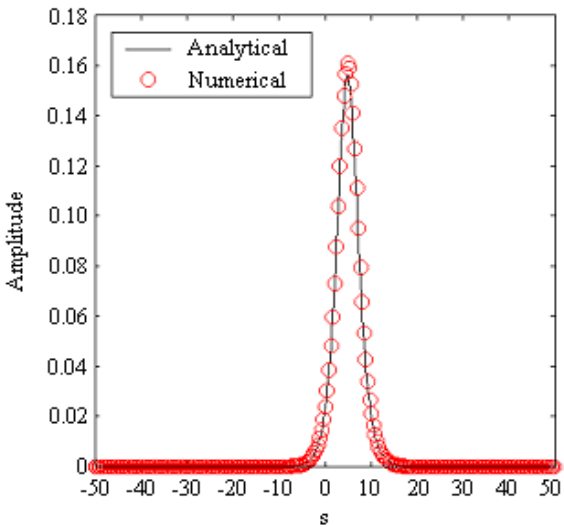

**Figure 9:** Analytical and numerical solutions for $|a_1(\xi,s)|$, in $\xi = 10$, for $\delta = -1/4$.

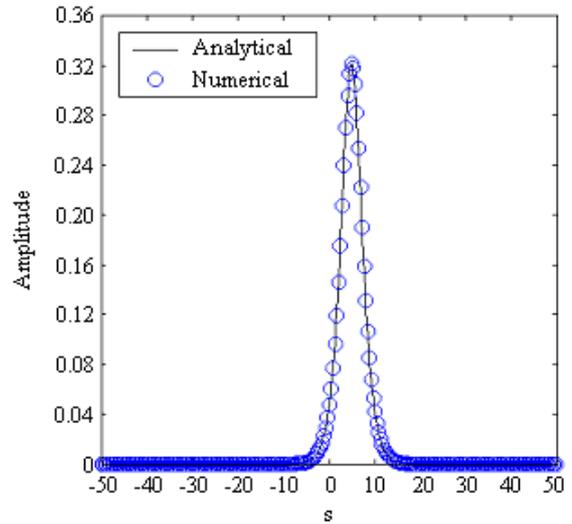

**Figure 10:** Analytical and numerical solutions for $|a_2(\xi,s)|$, in $\xi = 10$, for $\delta = -1/4$.

Figures 11 and 13 present the profiles of the fundamental harmonic and second harmonic modules, respectively, obtained numerically in all the computational domain, when $\delta = -1/4$. Figures 12 and 14 present the upper view of the propagation of such waves along the computational domain. In accordance with the analytical results (YMAI et al., 2004), from Figures 9 and 10, and from the upper views of the numerical solutions in all the computational domain, presented in figures 12 and 14, it is verified that the solitons waves, when $\delta \neq 0$, propagate in the spatial dimension $s$.

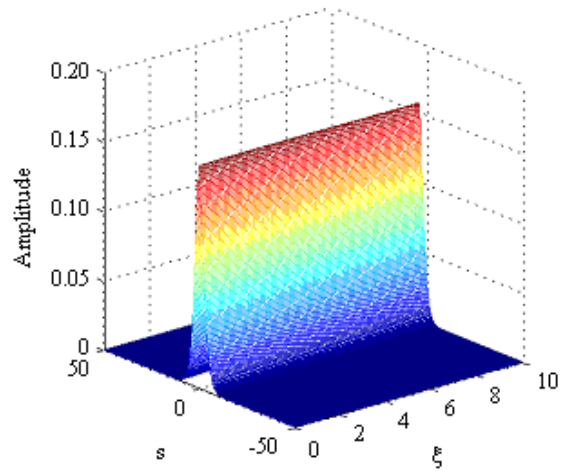

**Figure 11:** Numerical solution of $|a_1(\xi,s)|$, in all computational domain, for $\delta = -1/4$.

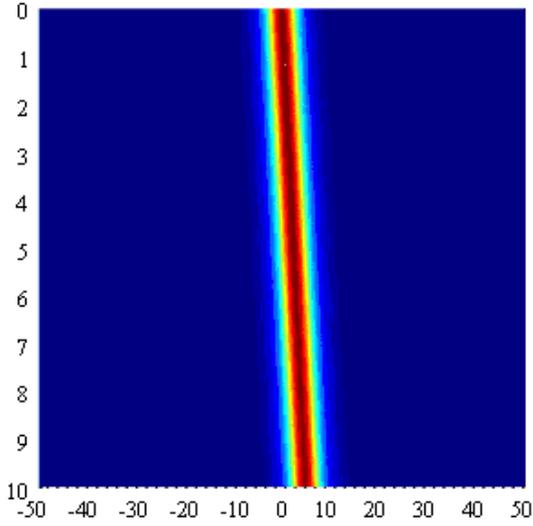

**Figure 12:** Upper view of $|a_1(\xi,s)|$ for $\delta = -1/4$.

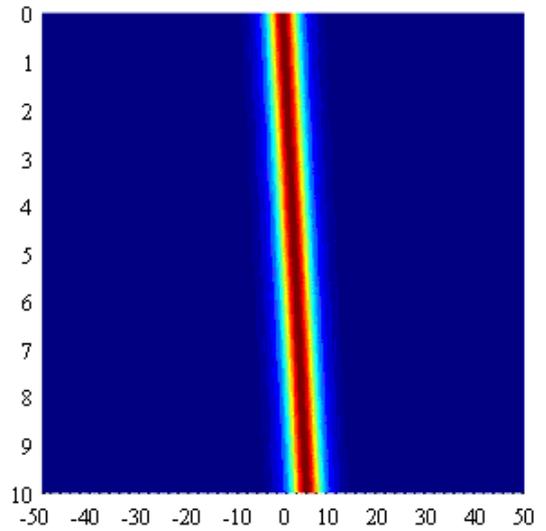

**Figure 14:** Upper view of $|a_2(\xi,s)|$ for $\delta = -1/4$.

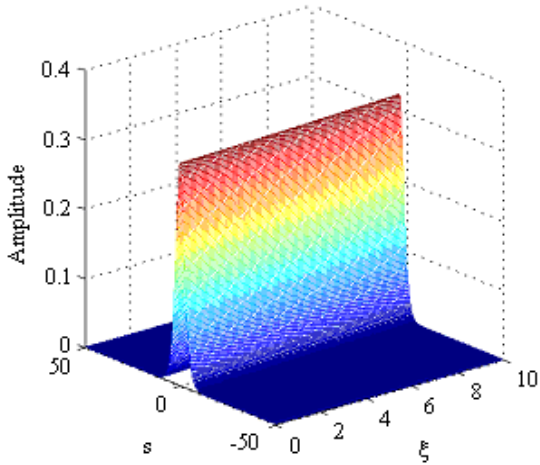

**Figure 13:** Numerical solution for $|a_2(\xi,s)|$, in all computational domain, for $\delta = -1/4$.

Notice that, in the analytical solutions (2-3), the velocity propagation $v$ of the solitons waves, in $s$, is given by

$$\left(s - \frac{r\delta}{2\alpha - r}\xi\right) = s - vt \implies v = \frac{r\delta}{2\alpha - r}, \quad (7)$$

where it is explicitly observed the dependence of $v$ on $\delta$. In the considered case, $\delta = -1/4 < 0$, with $r = -1$, $\beta = -1/2$, $\alpha = -1/4$, it is noticed that the soliton progressed in the positive direction of $s$, as expected.

## 5 Conclusions

The numerical scheme developed in this work, based on the finite difference method, was shown to be relatively simple from the computational-mathematical point of view, and adequate for the obtention of the numerical solitons solutions in ideal optical fibers.

It is the intention of the following works to describe the behavior of the quasi-soliton waves in non-ideal fibers, as well as to describe how the propagation and the stability of the soliton waves are affected when perturbatives processes are considered in the PDEs studied by YMAI et al. (2004). It should be observed that, at this level, analytical solutions are not known, so that the theoretical studies should be conducted by means of numerical procedures. As possible perturbatives processes that affect the propagation of solitons in dielectrical fibers, the following are mentioned:

1. absorptions of several types due to inhomogeneities, molecules of hydrogen and bubbles in the fiber (RAGHAVAN; AGRAWAL, 2000; BÖHM; MITSCHKE, 2007);
2. defects in the manufacture of the optical fiber like variations in the fiber diameter, rugosity, sinuosity in the longitudinal axis, micro curvatures and mends in the link by fusion with arc light (STROBEL, 2004; TSARAF; MALOMED, 2009);
3. noises in the electrical fields of the waves, for example, by the soliton pumping process, with the aim of compensating the absorption of the optical fiber (WERNER; DRUMMOND, 1993; LAU; KAHN, 2007);

4. high-order dispersion and high-order nonlinearities (AGRAWAL, 1995);
5. Raman effect and self-frequency shift (LATAS; FERREIRA, 2007);
6. polarization-mode dispersion (DRIBEN; MALOMED, 2007); among other disturbing processes.

In this context, the numerical procedure developed and validated in this work is intended to be used to approach the issue of propagation and stability of solitons in non-ideal optical fibers.

## Acknowledgements

The author P. L. Natti thanks the Universidade Estadual de Londrina for the financial support obtained by means of the Programs FAEPE/2005 and FAEPE/2009. The author C. F. de Oliveira thanks the Universidade Estadual de Londrina for the scholarships IC/UEL granted from August/2006 to July/2007 and from August/2008 to February/2009. The author N.M.L. Romeiro acknowledges CNPq-Brazil (National Council for Scientific and Technological Development) for the financial support to this research (CNPq 200118/2009-9).